\documentclass[showpacs,amsmath,amssymb,nofootinbib,prd]{revtex4}
\usepackage{graphicx}
\usepackage{enumerate}
\usepackage{amsmath}
\usepackage{amssymb}
\usepackage{amsfonts}
\usepackage{color}
\begin{document}
\title{Scalar-pseudoscalar interactions in neutrino-electron scattering}
\author{R. Gait\'an$^1$}
\email{rgaitan@unam.mx} 
\author{E. A. Garces$^2$}
\email{egarces@fis.cinvestav.mx} 
\author{O. G. Miranda$^2$}
\email{omr@fis.cinvestav.mx} 
\author{J.H. Montes de Oca Y.$^1$}
\email{josehalim@comunidad.unam.mx} 

\affiliation{$^1$Departamento de F\'isica, FES-Cuautitl\'an, UNAM, 
   Apdo. Postal 142 54700, Cuautitl\'an-Izcalli, Estado de M\'exico, 
   Mexico}
\affiliation{$^2$Departamento de F\'{\i}sica, Centro de
  Investigaci{\'o}n y de Estudios Avanzados del IPN\\ Apdo. Postal
  14-740 07000 Mexico, DF, Mexico}
\begin{abstract}
Many extensions to the Standard Model imply the existence of new
charged scalar Higgs bosons. 
We study the contribution of a general scalar or pseudoscalar coupling
for the neutrino-electron scattering.
We take a phenomenological approach in order to obtain model
independent limits to the couplings that arise in this picture.
We illustrate the reach of the constraints by studying the particular
case of the type III two Higgs doublet model, where we have found new
constraints to some elements of the Yukawa couplings mixing matrix
($|Y_{ee}| \leq 1 \times 10^{-1}$ 
and $|Y_{e\mu}| \leq 7\times10^{-2}$ at 90~\% CL).
\end{abstract}

\maketitle
\section{Introduction}
\label{sec-intro}
The observation of a new boson at a mass of 125~GeV reported by ATLAS
~\cite{atlas} and CMS~\cite{cms} collaborations maintains the
motivation to continue analyzing richer content in the scalar
sector. An interesting proposal arises when one or more scalar fields
are added to the scalar sector of the Standard Model (SM). One of the
simplest and interesting proposals adds an additional doublet scalar
field, named as Two Higgs Doublet Model (2HDM). As a result, new
charged Higgs bosons are predicted which could give a sign of physics
beyond the SM. This charged Higgs boson has been searched by ALEPH,
DELPHI, L3 and OPAL collaborations in charged Higgs boson
pair-production processes~\cite{lep}. Their results have excluded
values for charged Higgs boson mass below $72.5$~GeV or $80$~GeV at
95~\% C.L.  when the results are interpreted within 2HDM type I or
type II, respectively. The latest experimental searches for channels
$t \rightarrow H^\pm b$ and $H^\pm\rightarrow \tau \nu$ have been
carried out by ATLAS~\cite{atlas-h+} and CMS~\cite{cms-h+}. The CMS
collaboration reports upper limits on the branching fraction $Br(t
\rightarrow H^\pm b)$ in the range of 2-4~\% for charged Higgs boson
masses between 80 GeV and 160 GeV~\cite{cms}. Moreover, ATLAS
collaboration excluded the mixing angle $\beta$ of the charged scalar
in the region $2\leq \tan\beta\leq 6$ for charged Higgs
boson masses between 90~GeV and 150~GeV in the MSSM benchmark
scenario.

Other possible sources that could give clues to the existence of
charged Higgs boson are the electron and neutrino scattering
processes. The scattering among electrons and neutrinos plays an
important role in particle physics since it is a purely leptonic
process. As is well known, the Weinberg angle $\theta_W$ and the Fermi
constant $G_F$ are measured with high precision in these leptonic weak
interactions. Based on the SM, the only contributions for the $\nu-e$ or
$\bar{\nu}-e$ scattering are given by the $W^\pm$ or $Z^0$ gauge
bosons, which means purely weak interactions. The Higgs boson could
participate at tree level if the neutrinos are considered with masses
but its contribution is proportional to the neutrino masses which is
neglected. However a charged Higgs boson contribution to the process
could be considered without the neutrino masses assumption.

Beyond the SM framework, Non Standard Interactions (NSI) have been
considered with vector, axial or tensor structure to electron neutrino
scattering processes~\cite{garces2012,barranco2012}. Under this
framework, the scalar or pseudoscalar structure has not been
considered because its contribution is smaller. However, some authors
give a phenomenological study with the possibility of scalar or
pseudoscalar covariant neutral current interactions
\cite{kayser,kingsley,bardin}. 

In this paper we are focused on a general analysis of scalar and
pseudoscalar interactions that contribute to the
(anti)neutrino-electron scattering processes. The paper is organized
as follows: in Sec. II, we introduce the general formalism for the
scalar-pseudoscalar interactions and derive the differential cross
section for (anti)neutrino-electron scattering processes. In section
III we discuss briefly the experimental data that we will use to
obtain our constraints. Finally, in section IV we will show the
numerical results for the general formalism and we will discuss the
case of the 2HDM-type III.

\section{General cross section}
\label{sec-2}
In order to have a model independent scalar-pseudoscalar interaction
among leptons and charged Higgs bosons, we consider the effective
Lagrangian 
\begin{equation}
\mathcal{L}_{S,P} = \sum_{\alpha}\sum_{\beta}\overline{l}_{\alpha}\left(
\mathcal{O}_{S,P} \right)
\nu_{\beta},
\label{L_leptons-H-}
\end{equation}
where $\left(\mathcal{O}_{S,P} \right)$ is a general operator with
scalar and pseudoscalar interactions and
$\alpha,\beta=\,e,\,\mu,\,\tau$.  We will see in section~\ref{sec-3}
that this expression will be useful for some extensions of the
Standard Model. 

With this effective Lagrangian, the amplitude for the
$\nu_{\beta}-l_\alpha$ elastic scattering is given by the sum
\begin{equation}
\mathcal{M}_{\nu_\beta\,l_\alpha}=\mathcal{M}_{SM}+\mathcal{M}_{SP},
\end{equation}
with $\mathcal{M}_{SM}$ the usual amplitude of the Standard Model and
${M}_{SP}$ the new scalar-pseudoscalar contribution that, assuming a
low energy regime, is given by
\begin{equation}
\mathcal{M}_{SP}=\frac{-i}{M_{H}^{2}}\overline{l_\alpha}\left( p^{\prime }\right) \left(
\mathcal{O}_{S,P} \right) \nu_\beta\left( k\right) \overline{\nu_\beta}\left( k^{\prime
}\right) \left( \mathcal{O}_{S,P} \right) l_\alpha\left( p\right).
\end{equation}
It is important to note that the two interference terms
$\mathcal{M}_{SM}\mathcal{M}_{SP}^{\dag}$ and
$\mathcal{M}_{SP}\mathcal{M}_{SM}^{\dag}$ contained in
$\left|\mathcal{M}_{\nu_\beta\,l_\alpha}\right|^2$ give an important
contribution to the result. 
In order to calculate these terms it is
useful to make a Fierz reordering, using known
techniques~\cite{nieves-pal}. 
For the case of the operator $\mathcal{O}_{S,P}$, it is possible 
to find the general transformation 
\begin{eqnarray}
2\overline{l}_\alpha \left( a+b\gamma _{5}\right) \nu_\beta
\overline{\nu}_\beta \left( c+d\gamma _{5}\right) l_\alpha
\nonumber &=& \left( a+b\right) \left( c+d\right) \left( \overline{l}_{\alpha}P_{R}l_{\alpha}%
\overline{\nu}_{\beta}P_{R}\nu_{\beta}+\frac{1}{4}\overline{l}_{\alpha}\sigma ^{\mu \nu
}P_{R}l_{\alpha}\overline{\nu}_{\beta}\sigma _{\mu \nu }P_{R}\nu_{\beta}\right)   
\nonumber \\
&&+\left( a+b\right) \left( c-d\right) \overline{l}_{\alpha}\gamma ^{\mu
}P_{L}l_{\alpha}\overline{\nu}_{\beta}\gamma _{\mu }P_{R}\nu_{\beta}  \nonumber \\
&&+\left( a-b\right) \left( c+d\right) \overline{l}_{\alpha}\gamma ^{\mu
}P_{R}l_{\alpha}\overline{\nu}_{\beta}\gamma _{\mu }P_{L}\nu_{\beta}  \nonumber \\
&&+\left( a-b\right) \left( c-d\right) \left( \overline{l}_{\alpha}P_{L}l_{\alpha}\overline{\nu}_{\beta}P_{L}\nu_{\beta}+\frac{1}{4}\overline{l}%
_{\alpha}\sigma ^{\mu \nu }P_{L}l_{\alpha}\overline{\nu}_{\beta}\sigma _{\mu \nu
}P_{L}\nu_{\beta}\right),
\end{eqnarray}
where $a,\,b,\,c,\,d$ are constants and $P_{L,R}$ the usual chiral
projectors $\frac12(1\mp\gamma^5)$. 
Averaging over initial spin and summing over final
spin, we have
\begin{eqnarray}
\left\vert \overline{\mathcal{M}_{\nu_\beta\,l_\alpha}}\right\vert
^{2}&&=16G_{F}^{2}m_{e}^{2}E_{\nu}^{2}\left\{ \left[ \left( g_{A}-g_{V}\right)
^{2}+4\left( \left\vert g^{\alpha,\beta}_{S}\right\vert +\left\vert g^{\alpha,\beta}_{P}\right\vert \right) ^{2}+2\left( g_{V}-g_{A}\right) \textrm{Re}\left(
g^{\alpha,\beta}_{S}-g^{\alpha,\beta}_{P}\right) \right] \left( 1-y\right) ^{2}\right.
\nonumber \\
&&+\left. \left( g_{A}+g_{V}\right) ^{2}-\left[ \left(
g_{V}^{2}-g_{A}^{2}\right) +\left( g_{V}+g_{A}\right) \textrm{Re}%
\left( g^{\alpha,\beta}_{S}-g^{\alpha,\beta}_{P}\right) \right] \frac{m_{e}y}{E_{\nu}}%
\right\}   \label{Mp_f},
\end{eqnarray}
where the $g_{V,A}$ are the Standard Model values for the vector and axial coupling constants,  shown explicitly
in Table~\ref{table1}, 
and  $y=1-\frac{E_{\nu}^{\prime}}{E_{\nu}}$, for $E_{\nu}^{\prime}$
and $E_{\nu}$ the neutrino final and initial energies, respectively.
With the help of previous formula it can be easily obtained that the
neutrino-electron elastic cross section is
\begin{eqnarray}
\frac{d\sigma _{\nu_\beta\,l_\alpha}}{dT} &=&\frac{2G_{F}^{2}m_{e}}{\pi }\left\{
g_{L}^{2}+\left[ g_{R}^{2}+\left( \left\vert g^{\alpha,\beta}_{S}\right\vert
+\left\vert g^{\alpha,\beta}_{P}\right\vert \right) ^{2}+g_{R}\textrm{Re}\left(
g^{\alpha,\beta}_{S}-g^{\alpha,\beta}_{P}\right) \right] \left( 1-\frac{T}{E_{\nu}}%
\right) ^{2}\right.   \nonumber \\
&&\left. -\left[ g_{L}g_{R}+\frac{1}{2}g_{L}\textrm{Re}\left( g^{\alpha,\beta}_{S}-g^{\alpha,\beta}_{P}\right) \right] \frac{m_{e}T}{E_{\nu}^{2}}\right\},
\label{se_vl-e}
\end{eqnarray}
where $T$ is the electron recoil kinetic energy  and
$g_{L,R}=\frac{g_V\pm g_A}{2}$.  Analogously, the
antineutrino-electron elastic cross section can be obtained to be 
\begin{eqnarray}
\frac{d\sigma _{\bar{\nu}_\beta \,l_\alpha}}{dT} &=&\frac{2G_{F}^{2}m_{e}}{\pi }%
\left\{ \left[ g_{R}^{2}+\left( \left\vert g^{\alpha,\beta}_{S}\right\vert
+\left\vert g^{\alpha,\beta}_{P}\right\vert \right) ^{2}+g_{R}\textrm{Re}\left(
g^{\alpha,\beta}_{S}-g^{\alpha,\beta}_{P}\right) \right] +g_{L}^{2}\left( 1-\frac{T}{%
E_{\nu}}\right) ^{2}\right.   \nonumber \\
&&\left. -\left[ g_{L}g_{R}+\frac{1}{2}g_{L}\textrm{Re}\left( g^{\alpha,\beta}_{S}-g^{\alpha,\beta}_{P}\right) \right] \frac{m_{e}T}{E_{\nu}^{2}}\right\}.
\label{se_antiv-e}
\end{eqnarray}
In the following we will study the dimensionless
parameters $g^{\alpha,\beta}_{S,P}$ in muon and electron
(anti)neutrino scattering.  In general this parameters can be
different in each case.

\begin{table}
\centering
  \caption{Standard Model couplings for $\nu_i\,e$ elastic scattering. $g_{V,A}$ are the same for the $\bar{\nu}_e\, e$ elastic scattering. }
  \label{table1}
\begin{tabular}{|l|l|l|}
  \hline
  Reaction & $g_V$ & $g_A$ \\
  \hline
  $\nu_e\,e\rightarrow\nu_e\,e$ & $2\sin^2\theta_W+\frac{1}{2}$ & $\frac{1}{2}$ \\
  \hline
  $\nu_\mu\,e\rightarrow\nu_\mu\,e$ & $2\sin^2\theta_W-\frac{1}{2}$ & $-\frac{1}{2}$ \\
  \hline
\end{tabular}
\end{table}

\section{Elastic neutrino scattering experiments}
\label{sec-2.1}

The cross sections computed in the previous section can be used to
estimate the number of events expected in an (anti)neutrino electron
scattering experiment.  We will consider  experiments
that have measured the electron and muon neutrino scattering off
electrons, both for neutrino and antineutrino channels. 

The first experiment that we discuss in this section is the case of
the CsI(Tl) detector in the neutrino reactor experiment
TEXONO~\cite{Deniz:2009mu,Deniz:2010mp}. The experimental resolution
allows in this case a binned analysis of the data, the electron recoil
energy is divided into ten energy bins, from $3$ to $8$~MeV.  The
expected number of events for energy bin is given by the following
integral
\begin{equation}\label{eq:nth}
N_i ^{th}= \kappa \int^{T_{i+1}}_{T_i} \int_{E_\nu}  \left(\frac{d\sigma}{dT} \right) \lambda_{\phi}^{\bar{\nu}_e}(E_\nu)  dE_\nu dT ,
\end{equation}
here $\kappa$ is a factor that includes the number of electron targets
in the detector and the time exposure of the experiment, and
$\lambda_{\phi}^{\bar{\nu}_e}(E_\nu) $ is the antineutrino spectrum
parameterized as in reference~\cite{Mueller:2011nm} for the specific
radioactive isotope abundances in the Kuo Sheng 2.9 GW
reactor~\cite{Wong:2006nx}. We use for this analysis the differential
antineutrino electron scattering cross section
$\left(d\sigma/dT\right)$ shown in Eq.~(\ref{se_antiv-e}).

We can compute the theoretically expected number of events for a given
values of $g^{ee}_{s}$~and~$g^{ee}_{p}$ and perform a statistical
analysis by using a $\chi^2$ function defined by
\begin{equation}
\chi^2 = \sum_{i=1}^{10} \left( \frac{N_i^{exp} - N_i ^{th} }{\Delta_i^{stat}} \right)^2,
\end{equation}
where $N_i^{exp}$ is the event rate for the i$^{th}$ bin measured by
the experiment and $\Delta_i^{stat}$ is the associated statistical
uncertainty.

A similar analysis can be done for the case of the electron neutrino
scattering off electrons. In this case we can confront our theoretical
estimates with the experimental result of the LSND
experiment~\cite{Auerbach:2001wg}, that has reported the value of the
total cross section for electron neutrino scattering off electron as
$\sigma_{\nu_e e} = (10.1 \pm 1.1 \pm 1.0)\times E_{\nu_e} ({\rm
  MeV})\times 10^{-45} {\rm cm}^2$. In this case we add the statistical
and systematic errors in quadratures.

The same kind of statistical analysis can be done for the measurement
of the muon neutrino-electron cross section reported by the CHARM II
experiment~\cite{Vilain:1994qy}. As in the LSND case, in this
experiment the total cross section is reported in the whole energy
range, both for $\nu_\mu-e$ and $\bar{\nu_\mu}-e$ and, therefore, we
will have two bins in the $\chi^2$ function. 

\section{Results and discussion}
\label{sec-3}

\begin{figure}
\begin{minipage}{12pc}
\includegraphics[width=18pc]{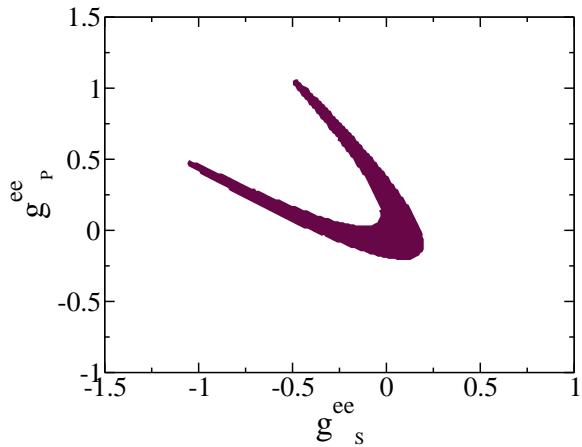}
\caption{\label{fig:2c} Parameter space allowed at 90~\% C.L. from 
the electron (anti)neutrino scattering off electrons. We have considered 
both the TEXONO and LSND results.}
\end{minipage}\hspace{3pc}%
\end{figure}

\begin{figure}
\begin{minipage}{12pc}
\includegraphics[width=18pc]{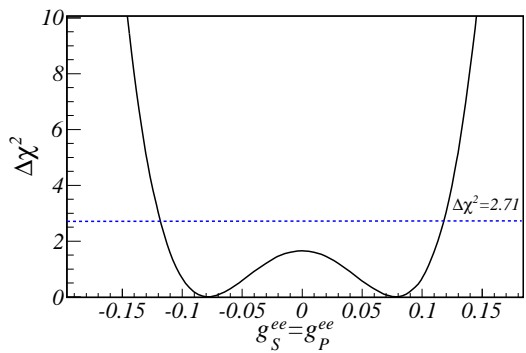}
\caption{\label{fig:3a} 
  Constraints from the electron (anti)neutrino
  scattering off electrons for the case $g^{ee}_P=g^{ee}_S$.
 We have
  considered both the TEXONO and LSND results.
  }
\end{minipage}\hspace{3pc}%
\end{figure}

In the previous sections we have showed the details of those 
computations in which are interested in. Here we show the results of our 
analysis and discuss our results. 

For the case of the electron (anti)neutrino scattering, we have
combined the results of the TEXONO and LSND analysis, described in the
previous section, in order to obtain constraints on the parameter
$g^{ee}_{S,P}$. The results are shown in Fig~(\ref{fig:2c}). Although
the allowed region appears to be large, it is important to notice that
this is due to the strong correlation between the scalar and
pseudoscalar couplings. Hence, despite the allowed region can have
values of $g_P^{ee}$ as large as $1$, this is only for the specific
case of $g_S^{ee}\approx -0.5$. We can see that the constraints for one
parameter at a time are more restrictive, given the regions $-0.15 \leq
g_P^{ee} \leq 0.35$ for $g_S^{ee} = 0$ and $-0.35 \leq g_S^{ee} \leq 0.15$
for $g_P^{ee} = 0$. As we will discuss below, another important case
will be that of $g_S^{ee} = g_P^{ee}$; the constraints for this
particular case are shown in Fig.~(\ref{fig:3a}) where it is possible
to see that $|g^{ee}_{s,p}|\le 0.12$ at 90~\% C.L.

\begin{figure}
\begin{minipage}{12pc}
\includegraphics[width=18pc]{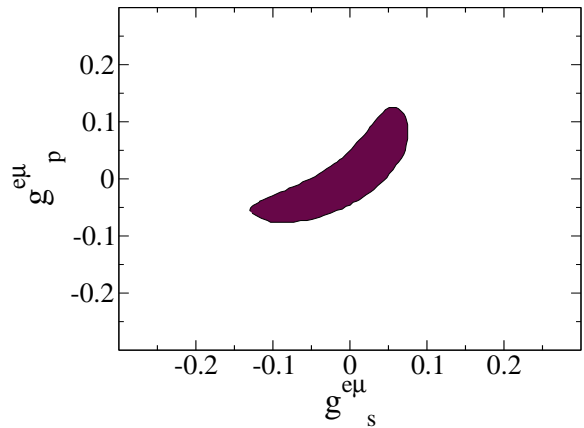}
\caption{\label{fig:3c} Parameter space allowed at 90~\% C.L. from the
  CHARM muon (anti)neutrino scattering data. }
\end{minipage}
\end{figure}

\begin{figure}
\begin{minipage}{12pc}
\includegraphics[width=18pc]{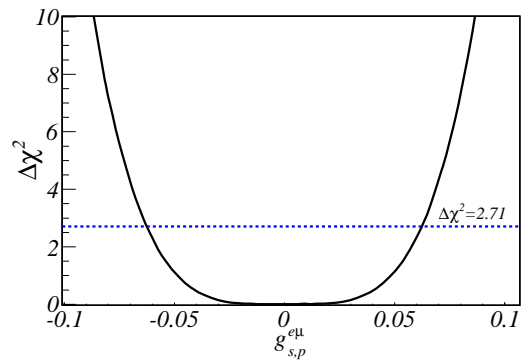}
\caption{\label{fig:3b} 
  Constraints from the muon (anti)neutrino
  scattering off electrons, for the case $g^{e\mu}_P=g^{e\mu}_S$,
  from the
  CHARM muon (anti)neutrino scattering data. }
\end{minipage}
\end{figure}

Similarly, results for the muon (anti)neutrino scattering can be
obtained by analyzing the CHARM II data. The results are shown in
Fig.~(\ref{fig:3c}), also at 90~\% C.L. As expected, this last case is
more restrictive thanks to the better statistics of the muon neutrino
experiments.  In particular, if one consider one parameter at a time
the constraint are stronger giving us the regions $-0.046 \leq
g_P^{e\mu} \leq 0.045 $ for $g_S^{e\mu} = 0$ and $ -0.032 \leq
g_S^{e\mu} \leq 0.036 $ for $g_P^{e\mu} = 0$. As in the previous
analysis we have also computed the fit for the case where $g_S^{ee} =
g_P^{ee}$ and we show this results in Fig.~(\ref{fig:3b}); in this
case the allowed values are $|g^{e\mu}_{s,p}|\le 0.06$ at 90~\% C. L.
 
As an application to a specific model we can consider the case of the 2HDM.
The 2HDM contains a charged Higgs boson which has interactions with leptons
as the introduced previously. In order to write explicitly the general
parameters $g_{S,P}$ we introduce the leptons and charged Higgs bosons
interactions in 2HDM. The most general structure of the Yukawa Lagrangian
for the leptons fields, can be written as follows:
\begin{equation}
\mathcal{L}_{Yukawa}^{leptons}=\sum_{a}\sum_{\alpha ,\beta }\overline{l}%
_{L_{\alpha }}Y_{a\alpha \beta }\phi _{a}e_{R_{\beta }}+h.c., 
\label{yukawa}
\end{equation}
where $\alpha ,\beta =e,\,\mu ,\,\tau $ and $a=1,2$ are indices for lepton
flavors and Higgs doublets, respectively. $l_{Li}$ denotes the left handed
leptons doublets and $e_{Rj}$ corresponds to the right handed singlets under
$SU(2)_{L}$. The Higgs doublets are decomposed as follows:
\begin{equation}
\phi _{a}=\left(
\begin{array}{c}
\varphi _{a}^{+} \\
\frac{v_{a}+\varphi _{a}+i\chi _{a}}{\sqrt{2}}%
\end{array}%
\right) ,  \label{doblete1}
\end{equation}%
where the vacuum expectation values $v_{a}$ are taken real and positive. One
of the $v_{a}$ could have a phase $e^{i\xi }$ in a more general case,
although this is outside the scope of our work. After getting a correct SSB
the charged leptons mass matrix is given as
\begin{equation}
M_{\alpha \beta }^{l}=\sum_{a}\frac{v_{a}}{\sqrt{2}}{Y}_{a\alpha \beta }
\label{m-y}
\end{equation}%
Note that Eq. (\ref{m-y}) relates the charged lepton mass matrix with the
two matrices $Y_{1}$ and $Y_{2}$. Therefore we can choose $Y_{1}$ as a
dependent variable. For the sake of simplicity from now on we will refer $%
Y_{2}=Y$. As usual, we introduce the $\beta $ angle in order to relate the
$%
\varphi _{a}^{\pm }$ with the physical $H^{\pm }$ and the Goldstone bosons
$%
G^{\pm }$. Then, the interactions among leptons and $H^{\pm }$ is given by
\begin{eqnarray}
\mathcal{L}_{H^{\pm }} &=&-\frac{\sqrt{2}}{v}\tan \beta H^{+}\overline{\nu
}%
_{\alpha }M_{\alpha \beta }^{l}P_{R}e_{\beta }+\frac{1}{\cos \beta }H^{+}%
\overline{\nu }_{\alpha }{Y}_{\alpha \beta }P_{R}e_{\beta }  \notag \\
&&-\frac{\sqrt{2}}{v}\tan \beta H^{-}\overline{e}_{\alpha }M_{\alpha \beta
}^{l}P_{L}\nu _{\beta }+\frac{1}{\cos \beta }H^{-}\overline{e}_{\alpha }{Y}%
_{\alpha \beta }^{\dag }P_{L}\nu _{\beta },  \label{H-2HDM}
\end{eqnarray}%
where $P_{L,R}=\frac{1\mp \gamma _{5}}{2}$. In literature, the 2HDM
types are defined through the structure of Yukawa
couplings~\cite{Branco:2011iw}. The general couplings given in
Eq. (\ref{yukawa}) denotes the 2HDM type III meanwhile the simplest
versions named as type II or type I are defined for $Y_{2}=0$ or
$Y_{1}=0$, respectively.

The specific cases of the 2HDM can be translated to the model
independent approach discussed here. The corresponding values for
$g_S$ and $g_P$ are shown in Table \ref{g_SP2HDM}. For these 
models the couplings are such that  $g_S=g_P$. 

Our model independent approach can be translated in the specific case
of 2HDM and the phenomenological bounds to the $g_{S,P}^{e\,\mu}$
parameters could give some insight in the parameter space of
$\tan^2\beta$ and $M_{H^-}$. The 2HDM would be the particular case
when $g_S=g_P$ and the table \ref{g_SP2HDM} shows
explicitly values for the different model types.

In this particular case we will
have $g_S=g_P$, and the values of these coupling constants are given
in Table \ref{g_SP2HDM}.

\begin{table}[tbp]
\caption{Explicit values for $g_{S,P}^{\protect\alpha\,\protect\beta}$
obtained for the two Higgs doublet models~\protect\cite{Branco:2011iw}.}
\label{g_SP2HDM}\centering
\begin{tabular}{|l|l|l|}
\hline
2HDM & $g_{S,P}^{e\,e}$ & $g_{S,P}^{e\,\mu}$ \\ \hline
type I & $\frac{\sqrt{2}m_{e}^{2}\cot ^{2}\beta }{v^{2}G_{F}M_{H}^{2}}$ & 0
\\ \hline
type II & $\frac{\sqrt{2}m_{e}^{2}\tan ^{2}\beta }{v^{2}G_{F}M_{H}^{2}}$ & 0
\\ \hline
type III & $\frac{1}{\sqrt{2}G_{F}M_{H}^{2}}\left(
\frac{\sqrt{2}m_{e}}{v}\tan \beta
-Y_{ee}\sec \beta \right) ^{2}$ & $\frac{Y_{e\mu }^{2}\sec ^{2}\beta
}{\sqrt{2}G_{F}M_{H}^{2}}$
 \\ \hline
\end{tabular}%
\end{table}

From the comparison of these constraints with the analytic expressions
shown in Table \ref{g_SP2HDM}, we can conclude that a constraint of
$|Y_{e\mu}|<7\times10^{-2}$ at 90~\% C.L. is reached when we consider a
charged Higgs mass of $100$~GeV~\cite{lep} and $\tan\beta = 1$ while
$|Y_{ee}|<0.10$ at 90~\% C.L., for the same values of the
Higgs mass and $\tan\beta$. These constraints are stronger than those
coming from lepton flavor violation in the charged leptonic
sector~\cite{Diaz:2004mk} where the following restrictions have been
found $|Y_{\mu e}|\leq 0.39$ and $|Y_{ee}|\leq 0.71$. We can conclude then
that, since we are dealing with tree level processes, we have been
able to obtain a stronger bound.  Moreover, the scattering that we are
studying is a two body process where the contribution to new physics
is only through charged current and, therefore there are less
parameters involved in the process (there is no dependence on
$\cos\alpha$ for example) and therefore the constraint is in some
sense more robust.

It may also be interesting to obtain similar constraints for different
textures of the mixing matrix since in this case constraints could be
more restrictive depending on the specific texture under
study~\cite{DiazCruz:2004tr,GomezBock:2009xz,Li:2010vf}.

\begin{acknowledgements}
This work has been supported by Conacyt grant 166639, by 
PAPIIT project IN117611-3, and by Sistema
Nacional de Investigadores (SNI),  M\'exico. J. H. M. O.
is thankful for the support from  postdoctoral DGAPA-UNAM grant.
\end{acknowledgements}


\end{document}